\documentclass[fleqn,10pt]{wlscirep}
\usepackage[utf8]{inputenc}
\usepackage[T1]{fontenc}
\usepackage{caption}
\usepackage{subcaption}
\usepackage{amsmath}
\usepackage{nccmath}
\usepackage{multicol}
\usepackage{ragged2e}

\usepackage{mwe} 
\usepackage{lipsum} 
\title{Nonlinear behavior of memristive devices for hardware
security primitives and neuromorphic computing systems}

\author[1,*]{Sahitya Yarragolla}
\author[1]{Torben Hemke}
\author[1]{Fares Jalled}
\author[2,3]{Tobias Gergs}
\author[2,3]{Jan Trieschmann}
\author[4]{Tolga Arul}
\author[1,$\dagger$]{Thomas Mussenbrock}

\affil[1]{Chair of Applied Electrodynamics and Plasma Technology, Ruhr University Bochum, Universitätsstraße 150, 44780 Bochum, Germany}
\affil[2]{Theoretical Electrical Engineering, Faculty of Engineering, Kiel University, Kaiserstraße 2, 24143 Kiel, Germany}
\affil[3]{Kiel Nano, Surface and Interface Science KiNSIS, Kiel University, Christian-Albrechts-Platz 4, 24118 Kiel, Germany}
\affil[4]{Chair of Reliable Distributed Systems, 
Faculty of Computer Science and Mathematics, 
University of Passau, D-94032 Passau, Germany}

\affil[*]{sahitya.yarragolla@rub.de}
\affil[$\dagger$]{thomas.mussenbrock@rub.de}

\begin{abstract}
Nonlinearity is a crucial characteristic for implementing hardware security primitives or neuromorphic computing systems. The main feature of all memristive devices is this nonlinear behavior observed in their current-voltage characteristics. To comprehend the nonlinear behavior, we have to understand the coexistence of resistive, capacitive, and inertia (virtual inductive) effects in these devices. These effects originate from corresponding physical and chemical processes in memristive devices. A physics-inspired compact model is employed to model and simulate interface-type RRAMs such as Au/BiFeO$_{3}$/Pt/Ti, Au/Nb$_{\rm x}$O$_{\rm y}$/Al$_{2}$O$_{3}$/Nb, while accounting for the modeling of capacitive and inertia effects. The simulated current-voltage characteristics align well with experimental data and accurately capture the non-zero crossing hysteresis generated by capacitive and inductive effects. This study examines the response of two devices to increasing frequencies, revealing a shift in their nonlinear behavior characterized by a reduced hysteresis range and increased chaotic behavior, as observed through internal state attractors. Fourier series analysis utilizing a sinusoidal input voltage of varying amplitudes and frequencies indicates harmonics or frequency components that considerably influence the functioning of RRAMs. Moreover, we propose and demonstrate the use of the frequency spectra as one of the fingerprints for memristive devices.

\end{abstract}
\begin{document}

\flushbottom
\maketitle
%
%
\thispagestyle{empty}

\section*{Introduction}

While CMOS devices have been the backbone of electronics for decades, memristive devices have emerged as a promising alternative with distinctive characteristics. Memristive devices exhibit intrinsic nonlinear behavior, as evidenced by the pinched hysteresis in their current-voltage characteristics (\textit{I}-\textit{V} curves), a well-established fingerprint defining their unique attributes~\cite{Yang2022}. These devices retain information about the amount of charge that has passed through them, presented in the history-dependent resistance function~\cite{Chua2011, Strukov2008, Waser2021}. As the charge accumulates, the resistance (conductance) of the memristor can alter, and this modification is retained until the charge is reset. 

Unlike the three traditional circuit components (resistor, capacitor, and inductor) that possess linear relationships between voltage and current, memristive devices display nonlinear behavior. The functioning of these devices relies on their internal state, which evolves over time and leads to intriguing and intricate nonlinear characteristics~\cite{Chua2019e}. The physical and chemical mechanisms that contribute to resistive switching, such as redox reactions, drift-diffusion of mobile defects, and filament formation~\cite{Waser2007} change the internal state of the device and thus the resistance of memristive devices. In nanoscale electronic devices, nonlinear behavior can arise from various unknown sources beyond those mentioned here. While memristive devices have prominent variable resistance, capacitive and virtual inductive effects can also contribute significantly to nonlinearity. To be more specific, the inductive effects are not real and are correlated and referred to as inertia effects, as described by Yarragolla et al.~\cite{Yarragolla2024, yarragolla2024b}. It can be manifested that resistive, capacitive, and inertia memory coexist in such devices~\cite{Qingjiang2014}. A detailed explanation is provided in subsequent sections.

The nonlinear behavior serves as the most important feature that is crucial for implementing neuromorphic dynamic systems such as memristive reservoirs, memristive oscillatory neural networks, and memristive chaotic systems~\cite{SETOUDEH2022, Yang2022}. Furthermore, the nonlinear behavior can be strategically utilized as a valuable entropy source for implementing hardware security applications such as physical unclonable functions (PUFs) or true random number generators (TRNGs)~\cite{Rajendran2021,Du2021}. By taking advantage of this behavior, the specific resistive states, transition patterns, harmonics, and resonant frequencies of memristive devices can be used as intrinsic and device-specific features. These characteristics, which are challenging to replicate or predict, can be utilized to create cryptographic keys, seeds, or true random numbers with high entropy. Although nonlinear behavior may be an important aspect, it presents several obstacles to overcome, such as achieving reproducibility and reliability in the face of unavoidable variability and ensuring stability and predictability in dynamic synaptic-like plasticity. These challenges require adequate investigation before pursuing practical and effective implementation of memristive devices. Therefore, comprehending the origin and examining the nonlinearity of memristive devices is imperative for their efficient utilization in neuromorphic computing and hardware security.  

           \begin{figure}[t]
                \centering             \includegraphics[width=0.67\textwidth]{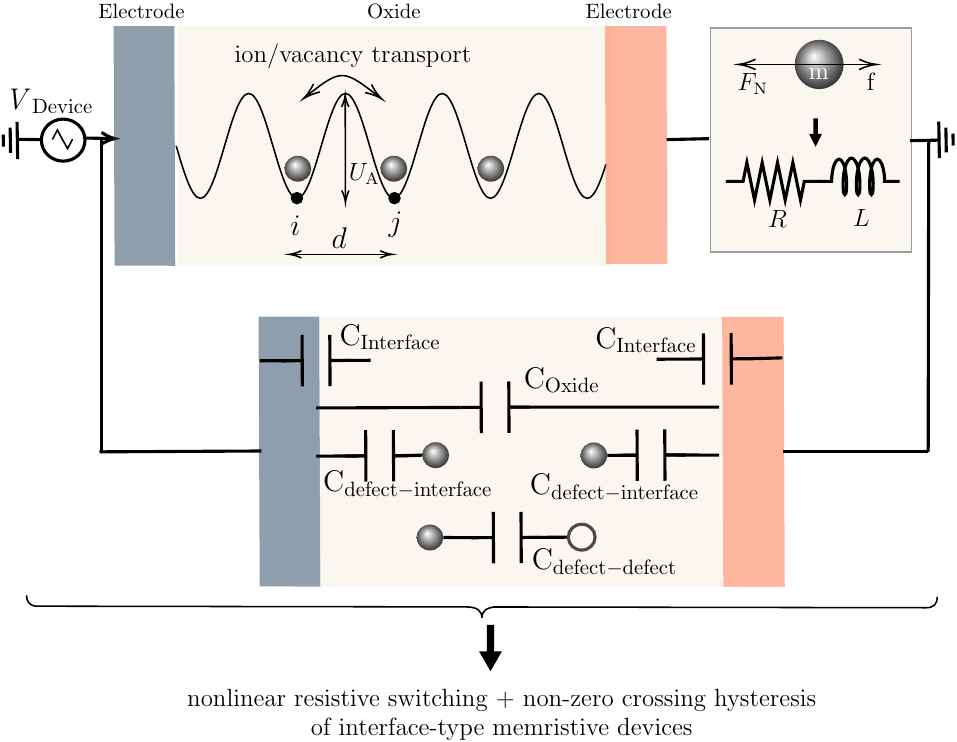}
                \caption{Illustration of the concert of co-existence of resistive, capacitive, and inductive effects in interface-type memristive devices.}
                \label{fig:1}
            \end{figure}

Memristive devices often exhibit a pinched hysteresis in their \textit{I}-\textit{V}  curves, with a zero crossing indicating zero current for zero input voltage. However, some devices, particularly RRAMs, display non-zero crossing hysteresis, suggesting the presence of stored charge. Several studies attribute this phenomenon to resistive, capacitive, and/or inertia effects~\cite{Qingjiang2014, Sun2020, Elbana2024}. So, a more appropriate approach would be to consider the change in resistive switching as a shift in memristive device impedance rather than just memristance. This modification reflects the interplay of different effects, which introduces a frequency-dependent impedance, resulting in nonlinearities and intricate responses to sinusoidal inputs.

Additional evidence supporting the existence of these effects is shown in Nyquist plots of their impedances, which display double semicircles \cite{Taibl2016, Bisquert2022, Marquardt2023}. One semicircle can be attributed to the presence of an RC circuit, while a negative curve could be interpreted as an inductive loop. It is important to note that this inductive behavior does not correspond to conventional electromagnetic inductance. Instead, we propose that it arises from inertial effects, which contribute to the formation of the second curve. The appearance of different semicircles in a Nyquist plot indicates the presence of multiple time constants or frequency-dependent behavior in a system. Therefore, the coexistence of resistive, capacitive, and inertia effects in memristive devices can result in diverse behaviors at distinct frequencies. Therefore, utilizing the frequency-dependent characteristics of memristive devices instead of the commonly used \textit{I}-\textit{V} curves can offer a different perspective on understanding the switching in these devices. 

From the previous discussion, it is critical to consider both particle (ion/vacancy) transport and capacitive effects when modeling memristive devices. These two components contribute significantly to the dynamic alterations in memristance. Capacitive effects, encompassing charge storage and redistribution within the device, are particularly imperative and play a pivotal role alongside ion transport mechanisms. The capacitive components of the device react as the charge accumulates or depletes, impacting the charge distribution and, subsequently, the resistance state. The dual contribution of ion transport and capacitive effects enhances the nonlinear properties observed in memristive devices. Understanding the capacitive contributions is critical to comprehending the complex dynamics that result in non-volatility, hysteresis, and memory-dependent changes in resistance. A deeper understanding of the nonlinear behavior of memristive devices can be achieved through comprehensive modeling of ion transport and capacitive effects. Inertia effects are not modeled independently in this work but are included through particle transport. Subsequent sections provide details on how inertia effects are integrated into the overall model for a comprehensive interpretation.

The paper examines the nonlinear behavior of interface-type memristive devices, utilizing a cloud-in-a-cell (CIC) scheme-based model for the double barrier memristive device (DBMD)~\cite{Yarragolla2022DBMD} and the bismuth ferrite oxide memristive device (BFO)~\cite{Yarragolla2022BFO, Yarragolla2024}. Unlike the state-of-the-art compact models~\cite{Kim2024, Bischoff2022, Thakkar2024}, this method more precisely incorporates the stochastic ion or vacancy transport like the multidimensional computational models~\cite{Abbaspour2020, Dirkmann2018, Kaniselvan2023, Aldana2023}, it is fast and accurate like the state-of-the-art compact models.  This study aims to develop a precise model for interface-type devices, considering capacitive effects, as the first step in understanding their nonlinear behavior. Therefore, to study the nonlinear behavior, the CIC models proposed by Yarragolla et al.~\cite{Yarragolla2022DBMD,Yarragolla2022BFO,Yarragolla2023} are further modified to incorporate the capacitive effects. The following section will provide a detailed explanation of the modeling of capacitive effects in interface-type memristive devices. Furthermore, this paper serves as a proof of concept for using the frequency spectra as a fingerprint to identify and characterize memristive devices, providing an alternative tool to the \textit{I}-\textit{V} curves for obtaining further insights into nonlinear dynamics. A parametric study is conducted to investigate the variations in resistive switching of the device resulting from various inputs and device parameters coupled with Fourier analysis of the frequency spectra, uncovering harmonics and sidebands.


           \begin{figure}[t]
                \centering
                \includegraphics[width=0.85\textwidth]{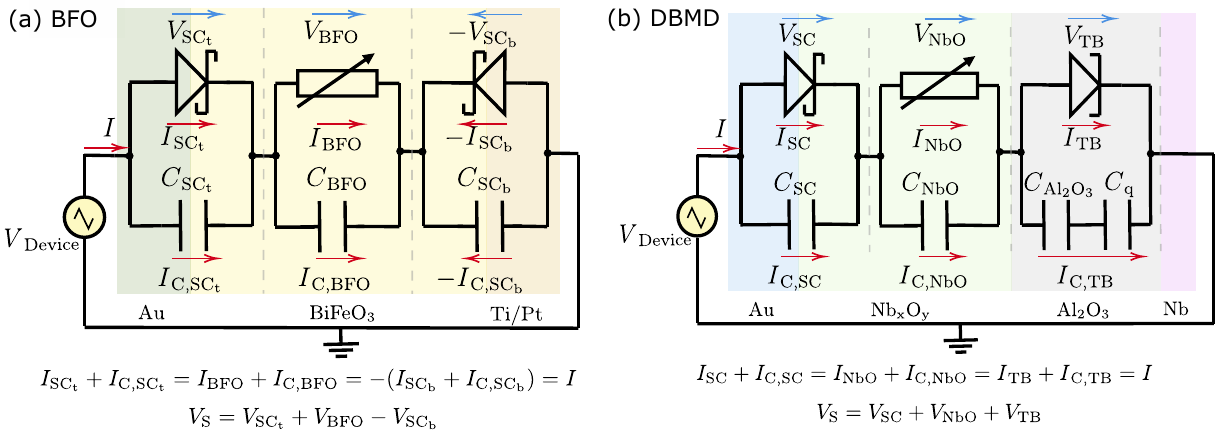}
                \caption{The modified equivalent circuits of (a) double barrier memristive device~\cite{Yarragolla2022DBMD} and (b) bismuth ferrite oxide memristive device~\cite{Yarragolla2022BFO} with parallel capacitors across different layers. SC: Schottky contact, TB: tunnel barrier, and SE: solid-state electrolyte.}
                \label{fig:2}
            \end{figure}

\section*{Simulation approach}

In the interface-type Au/BiFeO${3}$/Pt/Ti BFO ~\cite{Du2018} and Au/Nb$_{\rm x}$O$_{\rm y}$/Al${2}$O$_{3}$/Nb ~\cite{Hansen2015} devices, resistive switching occurs through particle transport, i.e., the drift-diffusion or trapping-de-trapping of charged defects, involving positively charged oxygen vacancies in BFO and negatively charged oxygen ions in DBMD. This mechanism alters the properties of the metal/oxide interface (Schottky or tunneling contacts), leading to an analog-type switching. The primary simulation models for DBMD and BFO devices, utilizing the CIC approach, have been detailed in ~\cite{Yarragolla2022DBMD} and ~\cite{Yarragolla2022BFO}, respectively.

The CIC method~\cite{Laux1996} is utilized to examine the behavior of particles in metal oxides, which contributes to resistive switching in RRAMs. This technique involves dividing the RRAM device into a lattice-like grid of cells, with each cell denoting a small volume within the device. Particles, representing atoms, ions, vacancies, or electrons, create a cloud around their positions, distributed across the grid with weights dependent on the oxide defect density. Weight represents the number of actual particles that each simulated particle denotes and allows for realistically modeling a larger population with fewer simulated particles. The weight is calculated as the ratio of total actual particles to simulated ones, which addresses computational limitations. For an input voltage bias, the particles are transported based on their drift velocity, calculated using the activation energy, ${U}_{\rm A}$, and electric field, $E$, obtained by solving the Poisson equation. The drift velocity is calculated as follows based on the probability of the particle movement from one lattice site to another ~\cite{bruce_1994, Meyer2008}:
\begin{ceqn}\begin{align} 
 v_{\rm D} = \nu_{0} d \,\, {\rm exp}\left ( -\frac{{U}_{\rm A}}{k_{\rm B}T} \right ) \sinh\left (\frac{\left | z \right |edE}{k_{\rm B}T}  \right ),
 \label{Eq:1}
\end{align} \end{ceqn}
\noindent where $d$ is the lattice constant, $\nu_{0}$ is the phonon frequency, $e$ is the elementary charge, $z$ is the charge number of the ion/vacancy, and $k_{\rm B} T$ is the product of Boltzmann constant and temperature. 

Once the ion or vacancy transport is completed, the electrical parameters, such as the currents and voltages across different layers, are calculated. To incorporate the capacitive effects, parallel capacitors are added across different layers in the equivalent circuit models of BFO and DBMD, as illustrated in Fig.~\ref{fig:2}. 

The resistive current through the Schottky contact is computed as follows ~\cite{Sze2007}:
\begin{ceqn}\begin{align} 
    I_{\rm SC} = A_{d}A^{*} T^{2}{\rm exp}\left \{ \frac{-\Phi_{\rm SC_{eff}}}{k_{B}T} \right \}\left ( {\rm exp}\left \{ \frac{eV_{\rm SC}}{n_{\rm SC_{eff}}k_{B}T} \right \} - 1\right ).
    \label{Eq:6}
\end{align} \end{ceqn}
\noindent Here $n_{\rm SC_{eff}}$ is the effective ideality factor, $\Phi_{\rm SC_{eff}}$ is the effective Schottky barrier height, and $A^{*}$ is the effective Richardson constant. The current across the tunneling barrier is calculated using the general Simmons current equation ~\cite{Simmons1963}.
\begin{ceqn}\begin{align} 
    I_{\rm TB} &= \frac{A_{d}e}{2\pi h\left (\beta d_{\rm TB_{eff}}  \right )^{2}}\Biggl( \Phi_{\rm TB_{eff}} \cdot {\rm exp}\left \{- {\rm A}\sqrt{\Phi_{\rm TB_{eff}}}\right \} - \left (\Phi_{\rm TB_{eff}}+ e\left | V_{\rm TB} \right |  \right ) \cdot {\rm exp}\left \{- {\rm A} \sqrt{\Phi_{\rm TB_{eff}}+ e\left | V_{\rm TB} \right | }\right \}\Biggr),
    \label{Eq:7}
\end{align} \end{ceqn}
\noindent where $\Phi_{\rm TB_{eff}}$ is the effective tunnel barrier height, $d_{\rm TB_{eff}}$ effective tunnel barrier width, $A_{d}$ is the device area and $A=\frac{4\pi\beta d\sqrt{2m} }{h}$. $\beta$, $m$, and $h$ are the correction factor, free electron mass, and the Planck constant. Moreover, the current across the oxide region is given by the general Ohm's law,
\begin{ceqn}\begin{align} 
    I_{\rm ox} = \sigma A_{\rm d}\frac{V_{\rm ox}}{l_{\rm ox}},
    \label{Eq:8}
\end{align} \end{ceqn}
\noindent where $l_{\rm SE}$ is the length of the active oxide layer and $\sigma$ is its conductivity. Here, 'ox' denotes either $\rm Nb_{\rm x}O_{\rm y}$ or $\rm BiFeO_{3}$ oxides. The variables with the subscript 'eff' represent their effective values, which are derived from the device's internal state, $q(t)$ and rate constant $\lambda$. For more information, refer to the detailed explanation provided by Yarragolla et al.~\cite{Yarragolla2022DBMD}. A flowchart of the proposed model (Supplementary Fig. S1), as well as a detailed explanation of the CIC approach (Supplementary Fig. S2) and Poisson solver, can be found in the supplementary material.

\begin{table}[!t]
  \centering
  \begin{tabular}{|l|c|c|c|}
    \hline
    \textbf{Component} & \textbf{depletion layer} & \textbf{capacitance} & \textbf{current} \\
    \hline
    \rule{0pt}{20pt} Schottky contact ~\cite{grundmann2015} & $d_{\rm{SC}}=\sqrt{\frac{2\epsilon_{0}\epsilon_{\textrm{r,ox}}\left ( \Phi_{\rm{SC_{eff}}}-eV_{\textrm{SC}}-k_{\textrm{B}}T \right )}{en}}$ & $C_{\textrm{SC}}= \frac{\epsilon_{0}\epsilon_{\textrm{r, ox}}A_{\textrm{d}}r_{\textrm{C}}}{d_{\rm{SC_{eff}}}}$ & $I_{\rm C,SC} =  C_{\rm SC}\frac{\mathrm{d} V_{\textrm{SC}}}{\mathrm{d} t}$  \\ 
    \rule{0pt}{20pt} & $d_{\rm{SC_{eff}}} = d_{\rm{SC}}(1 + \lambda_{d}\,q(t))$  &   &  \\ [2ex]
    \hline
    \rule{0pt}{20pt} Tunneling contact ~\cite{datta2005} & $d_{\rm{TB_{eff}}} = d_{\rm{TB}}(1 + \lambda_{t}\,q(t))$ & $C_{\textrm{ox,TB}}= \frac{\epsilon_{0}\epsilon_{\textrm{r,TB}}A_{\textrm{d}}r_{\textrm{C}}}{d_{\rm{TB_{eff}}}}$, \hspace{0.1cm} $C_{\textrm{q}}= e^{2}D(E)$ & $I_{\rm C,TB} =  C_{\rm TB}\frac{\mathrm{d} V_{\textrm{TB}}}{\mathrm{d} t}$ \\
    \rule{0pt}{20pt} & &  $C_{\textrm{TB}} = \frac{C_{\textrm{ox,TB}}C_{\textrm{q}}}{C_{\textrm{ox,TB}}+C_{\textrm{q}}}$ & \\ [2ex]
    \hline
    \rule{0pt}{20pt} Oxide & $d_{\rm{{eff}}} = \frac{\sum_{i=1}^{N_{\rm p}}\left ( \bar{x}_{\rm i}-\bar{x}_{\rm interface}\right )}{N_{\rm p}}$ & $C_{\textrm{ox}}= \frac{\epsilon_{0}\epsilon_{\textrm{r,ox}}A_{\textrm{d}}r_{\textrm{C}}}{d_{\rm{ox_{eff}}}}$ & $I_{\rm C,ox} =  C_{\rm ox}\frac{\mathrm{d} V_{\textrm{ox}}}{\mathrm{d} t}$ \\ [-1ex]
    \rule{0pt}{20pt}  & $d_{\rm{ox_{eff}}} = d_{\rm{eff_{mobile}}} - d_{\rm{eff_{fixed}}}$ &  & \\ [2ex]
    \hline
  \end{tabular}
  \caption{Equations used to compute capacitance across distinct layers in interface-type memristive devices.}
\label{table:1}
\end{table}

\subsection*{Capacitive effects}
The memristive device contains distinct capacitance elements that serve specific functions in different regions. The proposed model for an interface-type memristive device illustrates four categories of capacitive components as in Fig.~\ref{fig:1}: (a) capacitance between the interface and defects, (b) between positive and negative defects, (c) oxide capacitance, and (d) interface capacitance. These devices typically have Schottky or tunneling contacts as interfaces. The methods for modeling capacitance across these interfaces are detailed, and corresponding equations are provided in Table.~\ref{table:1}.
         
\textit{(a) Schottky Contact:} The capacitance of a Schottky contact is affected by the dynamics of the depletion layer, a region at the metal-semiconductor interface that lacks charge carriers. Variations in the effective width of the depletion layer $(d_{\rm SC_{eff}})$, caused by charge accumulation or depletion, affect the overall capacitance ~\cite{grundmann2015}. A positive voltage narrows the depletion layer, while a negative bias widens it, thus affecting the capacitance. The depletion layer's width can be determined by assessing charge accumulation due to traps at the interfaces between the metal and insulator ~\cite{Yan2013}. To determine the capacitance of the Schottky junction $(C_{\rm SC})$, we use the electrostatic capacitance equation for a parallel plate capacitor.

\textit{(b) Tunnel Barrier:} The capacitance across a tunnel barrier can be calculated by considering the series combination of oxide capacitance $(C_{\rm ox})$ and quantum capacitance $(C_{\rm q})$ ~\cite{John2004}. The oxide capacitance represents the traditional capacitive effects associated with the change in width of the oxide layer $(d_{\rm TB_{eff}})$, here in DBMD it is $\rm Al_{2}O_{3}$. The quantum capacitance accounts for the quantum mechanical effects related to the density of states ($D(E)$) in the semiconductor material. Quantum capacitance arises due to the changes in the density of states near the Fermi level in the semiconductor ~\cite{datta2005}. As the density of states varies, the quantum capacitance reflects the system's response to changes in charge distribution. The equation used for calculating the quantum capacitance and oxide capacitance in this work is given in Table~\ref{table:1}. The density of states is considered a constant term in this work for simplicity and is calculated based on the capacitance across the tunnel barrier measured by Hansen et al.\cite{Hansen2015}. Simulation methods such as density functional theory could be used to calculate it. Note that the quantum capacitance depends on the particular characteristics of the material, structure, and dimensions involved. Its contribution is more significant in two-dimensional (2D) materials and structures with prominent quantum confinement effects. For very thin layers like a tunnel barrier with a thickness of less than 2 nm, quantum effects become essential. 

\textit{(a) Metal Oxide:} The standard electrostatic capacitance equation is used to determine capacitance across a solid-state electrolyte. This equation considers the effective capacitance resulting from interactions between charged defects and the interface and among different charged defects. The depletion layer width, $d_{\rm ox_{eff}}$, indicates the difference in the average relative distance between the positions of mobile defects $d_{\rm eff_{mobile}}$ and the average relative distance between the positions of fixed defects $d_{\rm eff_{fixed}}$. When oxygen vacancies drift toward the Pt electrode in the BFO device or oxygen ions drift toward the Au electrode in the DBMD device for a positive bias, charge separation occurs within the oxide layer. This creates a structure similar to a capacitor where negatively charged defects are separated from positively charged defects. In our simulation model, we include fixed defects to maintain charge neutrality.     


Capacitance is typically multiplied by the roughness factor, a dimensionless quantity that indicates the deviation of the real surface area from the ideal geometric surface area. Similarly, a correction factor $(r_{\rm C})$ is applied to the electrostatic capacitance of various RRAM layers to adjust simulation results with experimental findings. This correction factor serves to compensate for several non-idealities inherent to nanoscale devices. Factors such as heterogeneous dielectrics within the oxide layer, interface effects between the oxide and adjacent materials, capacitance frequency dependency, variations in oxide thickness, quantum effects at the nanoscale, and process-related variability can collectively affect electrostatic capacitance. The value of $r_{\rm C}$, could be any number between 0 and 1, signifying a reduction in capacitance, accounting for non-idealities and deviations from theoretical expectations. By iteratively adjusting the correction factor, a best-fit value is sought that minimizes the discrepancy between theoretical and experimental results across different conditions and configurations.

\subsection*{Inductive effects}
A previously published meminductance model, which is controlled by flux or current and dependent on the system state, has been established in the literature~\cite{Ventra2009,Biolek2011}. Although Qingjiang et al.~\cite{Qingjiang2014} showed various effects in memristive devices, their use of the meminductance model may not be practical or accurate in simulating real-world devices. The assertion of electromagnetic inductance in nanoscale devices may not align with physical constraints, particularly when related to the state equation and a change in inductance proportional to the square of the number of turns. The term 'number of turns' in this context refers to internal parameters or states rather than the physical coil windings in a traditional inductive coil. The existence of an additional semi-circle in a Nyquist plot with negative impedance values is attributed to inductive or negative capacitive effects. Marquardt et al.~\cite{Marquardt2023} noted that physically realizing windings of an inductance of the required size (>10 mH) within a multi-layer system of approximately 10 nm thickness becomes impractical. However, Bisquert et al.~\cite{Bisquert2022} recently proposed a chemical inductor model that is non-electromagnetic, using an inductor and resistor to emulate these effects. While acknowledging the viability of an inductor and resistor model, we propose that the observed ``inductive effects'' are physically attributed to the inertia of charged particles (including ions, vacancies, electrons, or holes). Particle transport appears to be the most likely cause of the effect, although other unknown processes may also play a role.  As explained by Yarragolla et al.~\cite{Yarragolla2024}, this perspective aligns with the understanding that charged particles exhibit reluctance to a change in motion due to inertia, contributing to the observed ``inductive'' behavior in memristive devices. As induction is always coupled to a temporal change magnetic flux, the terminology ``inductive'' may be misleading. Ultimately, there is no magnetic field at all in the RRAMs under study. 

Furthermore, the kinetic inductance of superconducting linear structures is related to the inertia of free electrons \cite{Meservey1969,Annunziata2010}. The kinetic inductance is a natural derivation from the Drude model of electrical conduction. The Drude model of electrical conduction can be adapted to explain the motion of ions or vacancies in solid-state physics. Originally developed to describe the behavior of free electrons in a conductor, the Drude model can also be applied to RRAMs by considering the principles of momentum conservation and the response of charged particles to electric fields. Therefore, the kinetic inductance in an n-type $\rm Nb_{\rm x}O_{\rm y}$ layer in the DBMD device and an n-type $\rm BiFeO_{3}$ layer in the BFO device could be related to the inertia of the free electrons and the charged defects (ions and vacancies). A mathematical explanation of this relationship is given below.

In solid-state physics and electronic devices, the behavior of charged particles undergoing resistive switching mimics that of electrical circuits. The movement of these particles, considering drift velocity and friction, is encapsulated in a momentum conservation equation, notably expressed as $m\frac{\mathrm{d} v_{\mathrm{D}}}{\mathrm{d} t}= e E - m\gamma v_{\mathrm{D}}$. This equation, when extended to the drift velocity equation and then to a generalized Ohm's law ($\frac{\mathrm{d} j}{\mathrm{d} t}= \frac{e^2 n}{m} E - \gamma j$), forms the foundation of the Drude model for electrical conduction. Transitioning further, the equation for drift velocity, when applied to a one-dimensional scenario, yields the final expression~\cite{Yarragolla2024, yarragolla2024b} 
\begin{ceqn}\begin{align} 
        V_{\rm ox}= \underset{\textrm{kinetic inductance}}{\underbrace{\frac{m l_{\rm ox}}{e^2 n A_{\rm d}}}}\frac{\mathrm{d} I_{\rm ox}}{\mathrm{d} t} +\underset{\textrm{electrical resistance}}{\underbrace{ \frac{m l_{\rm ox}}{e^2 n A_{\rm d}} \gamma }}I_{\rm ox} = L_{\rm kinetic, ox} \frac{\mathrm{d} I_{\rm ox}}{\mathrm{d} t} + R_{\rm ox} I_{\rm ox},
    \label{Eq:9}
\end{align} \end{ceqn}
where $L_{\rm ox}$ is the kinetic inductance and $R_{\rm ox}$ is the electrical resistance of the oxide layer. To explain inertia effects more clearly, when the electric field and velocity are in phase, the charged particle responds promptly to changes in the electric field due to the absence of inertia. However, when they are out of phase, inertia causes a delay in the response of the charged particle to changes in the electric field, leading to a phase difference between them. Due to inertia, the velocity of particles and electric field will have a phase difference, similar to the inductive effect, where inductive current lags behind the voltage across the inductor due to induced electromotive force.


In our model, we account for inductance, referred to as the inertia effects in this paper, by considering the movement of charged particles related to their drift velocity in Eq.~\eqref{Eq:1}. This movement is influenced by the electric field, activation energy for migration, trapping, or de-trapping of ions or vacancies, as well as the probability of jumps. The equation for drift velocity effectively describes the average motion of ions and vacancies in response to an electric field, incorporating momentum-related inertia effects (virtual or inductance). This integrated approach streamlines the model, eliminating the need for a separate evaluation while explaining the intricate motion of charged particles within an oxide.

\begin{figure}[!t]
\centering\includegraphics[width=0.99\textwidth]{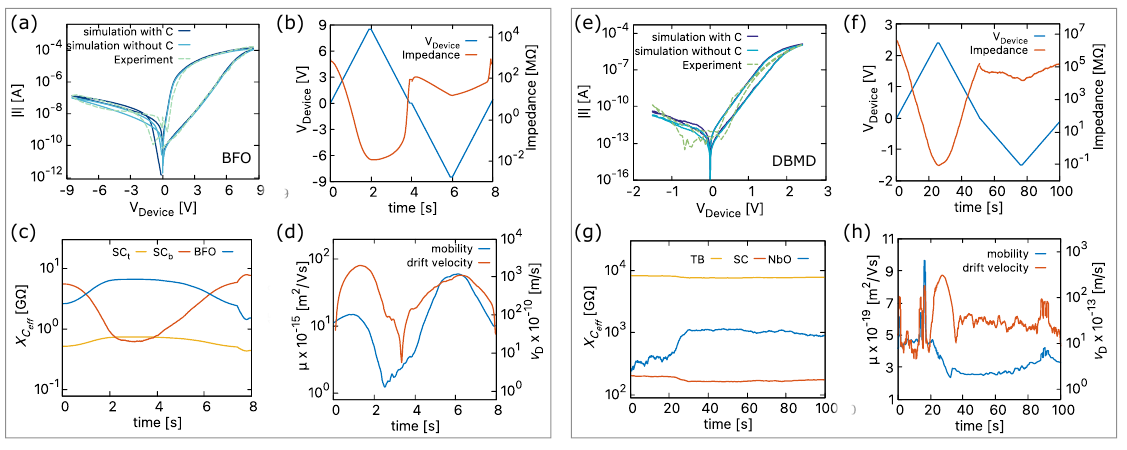}
\caption{The interface-type switching in (a)-(d) BFO and (e)-(h) DBMD devices, including capacitive and inertia effects. (a),(e) The simulated \textit{I}-\textit{V} curves with and without the capacitive effects.; (b),(f) the source voltage and change in impedance; (c),(g) the change in capacitive reactance and (d), (h) the absolute average mobility and absolute drift velocity of all mobile oxygen vacancies in BFO and mobile oxygen ions in DBMD memristive devices.}
\label{fig:3}
\end{figure}




\section*{Results and discussion}

The \textit{I}-\textit{V} curves shown in Fig.~\ref{fig:3} illustrate the nonlinear behavior of BFO and DBMD devices resulting from the coexistence of resistive, capacitive, and inertia effects. To obtain the \textit{I}-\textit{V} curves of BFO device shown in Fig.~\ref{fig:3}(a), an input voltage of 8.5\,V was utilized, in conjunction with the parameters used by Yarragolla et al.~\cite{Yarragolla2022BFO} (given in Supplementary Table. S1). The figure compares the \textit{I}-\textit{V} curves of BFO with and without including capacitive effects. Although the curves with and without capacitive effects exhibit a similar nonlinear change in current with the input voltage, the curve with included capacitive effects highlights a noticeable non-zero crossing hysteresis. This curve depicts the precise analog change in current with the crossing point observed at approximately -2.9\,V in the experimental \textit{I}-\textit{V} curves. This confirms the presence of charges in interfacial-type memristive devices and how they result in capacitive effects. 

In contrast to the BFO device, the \textit{I}-\textit{V} curves of the DBMD device, as shown in Fig.~\ref{fig:3}(e), exhibit a notable absence of non-zero crossing hysteresis. This difference can be attributed to the much higher capacitive reactance in the DBMD device as seen in Fig.~\ref{fig:3}(g), which is at least two orders of magnitude higher than that calculated for the BFO device as in Fig.~\ref{fig:3}(c). The simulated \textit{I}-\textit{V} curves for the DBMD obtained for parameters given in Supplementary Table. S2, both with and without the inclusion of capacitive effects, agree remarkably well with the experimental \textit{I}-\textit{V} curves. The reduced capacitance in the pF range contributes to the subtle variations observed in the curves with and without capacitive effects, indicating a minimal effect on the current. Notably, at lower input voltage frequencies for both BFO and DBMD, the inclusion or exclusion of capacitive effects may not result in significant changes in current. However, as the input voltage frequency increases, the capacitive effects become more pronounced, potentially affecting the overall device behavior (detailed later). To more effectively study the resistive switching mechanisms in these devices, we examined the impedance changes over time, as shown in Fig.~\ref{fig:3}(b) and Fig.~\ref{fig:3}(f). We opted to plot impedance to account for resistive, capacitive, and inertial effects, as opposed to the conventional approach of plotting resistance. As the voltage increases, vacancies in the BFO device and ions in DBMD migrate towards the Pt and Au electrodes, respectively, transitioning the device from a high impedance state (HIS) to a low impedance state (LIS) and vice versa.

            \begin{figure}[!t]
                \centering      \includegraphics[width=0.99\textwidth]{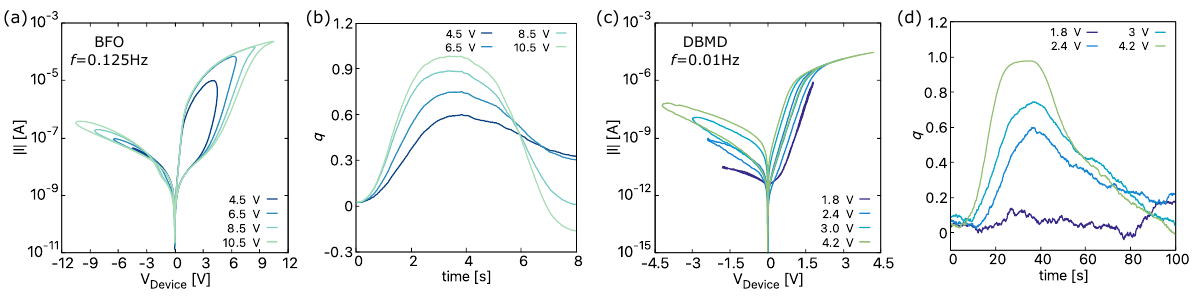}
                \caption{The \textit{I}-\textit{V} of (a) BFO and (c) DBMD devices for sinusoidal voltage of constant frequency and varying maximum voltage amplitude and their corresponding change in internal state shown in (b) and (d).}
                \label{fig:7}
            \end{figure}

We analyze the particle transport by plotting the average mobility and drift velocity of vacancies or ions over time, as shown in Fig.~\ref{fig:3}(d) and Fig.~\ref{fig:3}(h). During the part of the cycle with increasing positive voltage, the particle velocity increases until the activation energy is relatively low, while the mobility remains moderate. For a decreasing voltage from 8.5\,V to 0\,V, the activation energy is more near the Pt electrode, so the drift velocity and the mobility are reduced. Conversely, for negative voltage, the mobility of vacancies increases by two orders of magnitude, as observed by Du et al~\cite{Du2018}. However, in DBMD, the mobility of ions does not increase significantly, and the velocity remains relatively stable. The activation energy for oxygen vacancy drift in BFO ranges from approximately 0.55\,eV at the Au electrode to 0.76\,eV near the Pt electrode due to the diffusion of Ti$^{4+}$ ions into BiFeO$_{3}$ layer \cite{Yarragolla2022BFO}. Meanwhile, for oxygen ion drift in DBMD, it remains nearly constant at 0.76 eV across the $\rm Nb_{\rm x}O_{\rm y}$ layer \cite{Yarragolla2022DBMD}. As a result, the velocity of ions in DBMD is significantly lower than that of vacancies in BFO, leading to lower mobility in DBMD compared to BFO. 

\begin{figure}[!t]
                \centering      \includegraphics[width=0.99\textwidth]{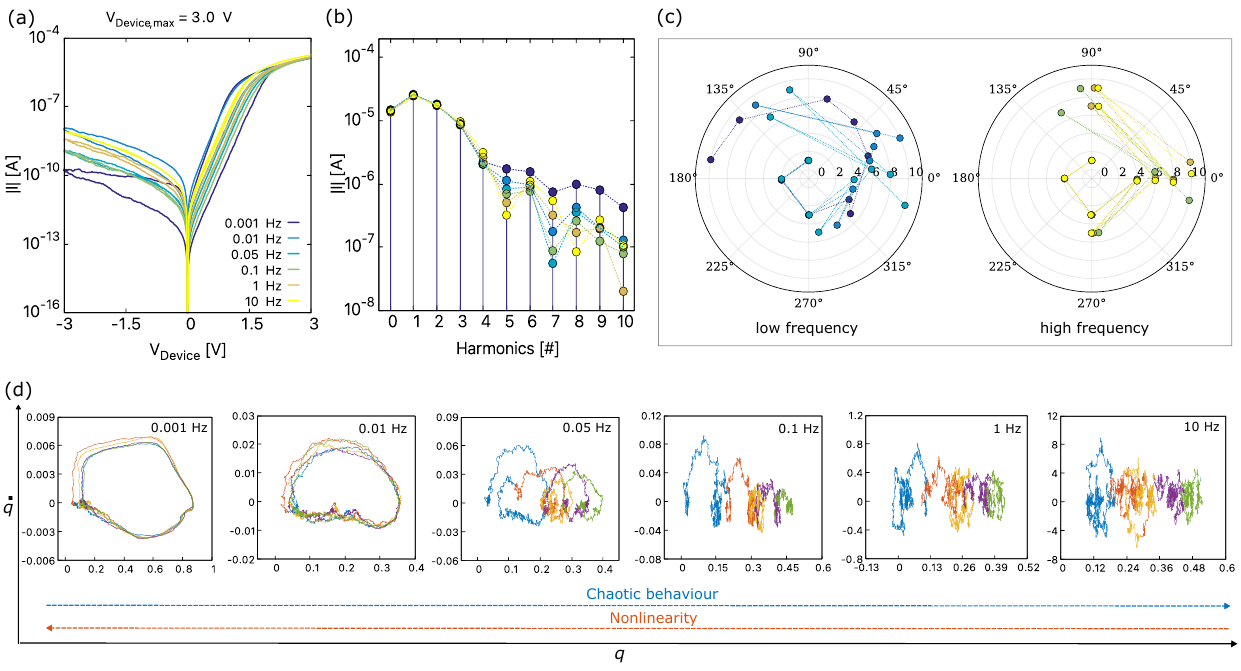}
                \caption{(a) The simulated \textit{I}-\textit{V} curves of DBMD obtained for a sinusoidal input voltage of 3\,V with various frequencies, Frequency spectra of DBMD corresponding to the \textit{I}-\textit{V} curves depicted in Figure (a) showing (b) amplitude spectra and (c) phase spectra, and (d) the attractors depict the internal state change of the device for different frequencies, plotted for five consecutive input voltage cycles. The legend for figure (a) applies to figures (b) and (c). Points in Figures (a) and (b) that are not immediately visible are superimposed.}
                \label{fig:5}
            \end{figure}

The nonlinearity in memristive devices undergoes significant changes with variations in frequency; therefore, understanding the frequency-dependent behavior of such devices is crucial. For this, different \textit{I}-\textit{V} curves of both devices are plotted in Fig.~\ref{fig:7} and Fig.~\ref{fig:5} for a sinusoidal input voltage of different amplitudes and frequencies. The following observations can be drawn from the three plots: 

\noindent\textbf{(a)} For a constant frequency and an increasing maximum device voltage $(V_{\rm Device, max})$, the hysteresis lobe area increases for BFO shown in Fig.~\ref{fig:7}(a) and DBMD in Fig.~\ref{fig:7}(c)~\cite{Izquierdo2021, Yarragolla2022DBMD}. This is because the device's internal state, measured based on the position of oxygen vacancies in BFO (Fig.~\ref{fig:7}(b)) and oxygen ions in DBMD Fig.~\ref{fig:7}(d), undergoes greater changes at higher voltages. As the voltage increases (decreases), the charged particles experience a greater (lesser) force, resulting in their drift toward the respective interfaces. This leads to an increase (decrease) in current and the difference between the low and high resistance states. However, this argument holds true only for the lower frequency range, a common feature observed in most memristive devices.

\begin{figure}[!t]
                \centering      \includegraphics[width=0.99\textwidth]{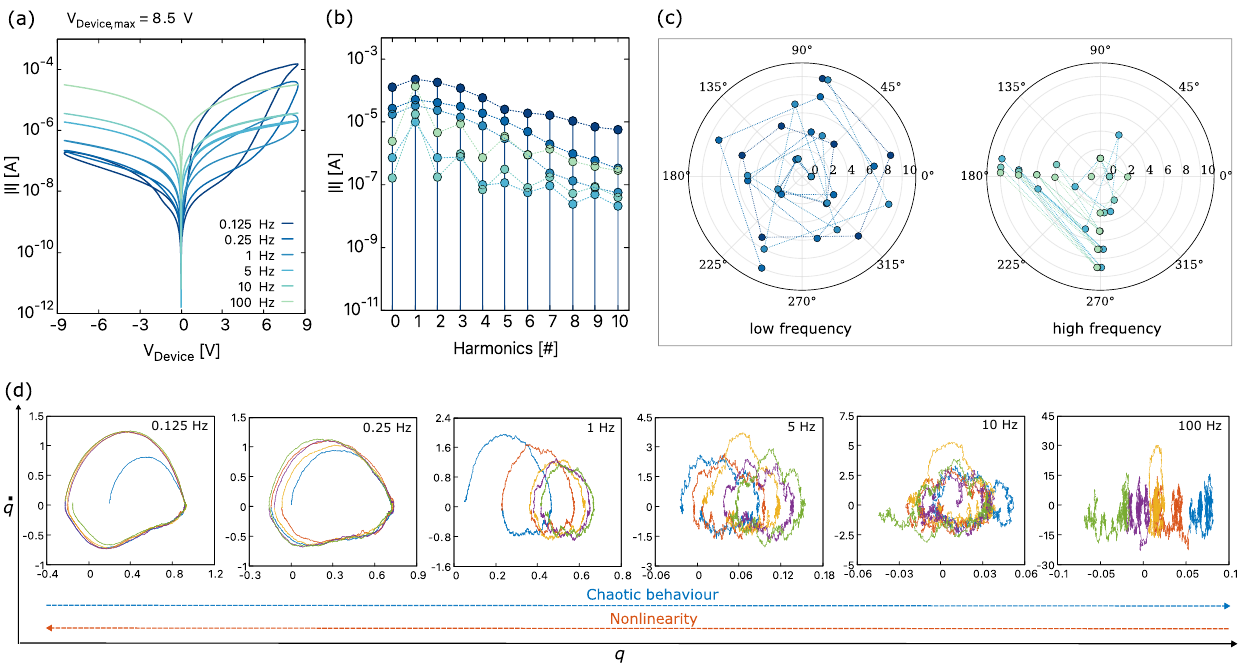}
                \caption{(a) The simulated \textit{I}-\textit{V} curves of BFO obtained for a sinusoidal input voltage of 8.5\,V with various frequencies, Frequency spectra of BFO corresponding to the \textit{I}-\textit{V} curves depicted in Figure (a) showing (b) amplitude spectra and (c) phase spectra, (d) the attractors depict the internal state change of the device for different frequencies, plotted for five consecutive input voltage cycles. The legend for figure (a) applies to figures (b) and (c). Points in Figures (a) and (b) that are not immediately visible are superimposed.}
                \label{fig:4}
            \end{figure}

\noindent\textbf{(b)} For a constant $(V_{\rm Device, max})$, as the frequency increases, the hysteresis loop area decreases as depicted in Figs.~\ref{fig:4}(a) and \ref{fig:5}(a) ~\cite{Izquierdo2021, Yarragolla2022DBMD}. This phenomenon occurs because the vacancies lack time to jump to a different lattice position and become confined in the same potential well. At lower frequencies, the slower response of the device allows more time for resistive switching mechanisms to manifest, resulting in a larger hysteresis area. As the frequency exceeds 10\,Hz for BFO and 1\,Hz for DBMD, the  \textit{I}-\textit{V} curves begin to exhibit behavior similar to that of a normal resistor. Moreover, the change in hysteresis area with frequency in memristive devices can be attributed to the dynamic interplay of capacitive and resistive effects. As the frequency increases, the capacitive reactance $(X_{\rm C}=1/2\pi fC)$ decreases, the inductive reactance $(X_{\rm L}=2\pi fL)$ increases, and vice versa. The impedance equation $Z=R+j(X_{L}-X_{C})$ illustrates that capacitive and inductive reactances balance at certain frequencies, approaching a more resistive behavior. The behavior of memristive circuits changes from nonlinear to more or less linear resistor-like behavior as the operational frequency varies due to the varying contribution of resistive, capacitive, and inertia effects.

\noindent\textbf{(c)} The presence of different effects in these devices has a substantial impact on the zero-crossing current in memristive devices when the frequency of the applied voltage changes. Fig.~\ref{fig:4}(a) shows that the crossing points move further from the origin as the frequency increases ~\cite{Qingjiang2014}, and this effect becomes more pronounced with increasing voltage. As frequency increases, the capacitive current also increases due to the reduced capacitive reactance. This observation was previously made by Yarragolla et al. \cite{Yarragolla2024}. The battery effect in a circuit becomes more pronounced with increasing frequency. The battery effect, commonly linked to the existence of capacitive components, is characterized by the storage and release of charges even when there is no external voltage source. The increased capacitance with frequency is accompanied by a decrease in capacitive reactance, which improves the capacity to store charges. The observed increased capacitance leads to a more pronounced battery-like effect, where the capacitive component functions as a virtual battery, storing charges and impacting the circuit's overall behavior. As a result, this is associated with a shift of the crossing point further away from the origin. Moreover, the effects of inertia also contribute to the shift in the crossing points. This has been explained in detail by Yarragolla et al. ~\cite{Yarragolla2024}. For a frequency of 10\, Hz, the change in current is almost linear, and the non-zero crossing disappears.

To evaluate the presence of harmonics, we present the frequency spectra of the BFO and DBMD devices in Fig.~\ref{fig:4}(b) and Fig.~\ref{fig:5}(b), respectively, corresponding to the \textit{I}-\textit{V} curves shown in Fig.~\ref{fig:4}(a) and Fig.~\ref{fig:5}(a). Both devices exhibit a consistent pattern in their amplitude spectra. At low frequencies, we observe a gradual reduction in spikes, indicating a decrease in the dominance of harmonic components in the signal. This phenomenon can be explained by the fact that the impedance obtained from resistive, capacitive, and inertial components exhibits less variation or better balance at lower frequencies. In other words, a uniform mixing of both even and odd harmonics is present at low frequencies leading to a uniform convergence of the amplitude spectrum. 
It is important to note that no resonances were observed in BFO and DBMD. However, it is possible that resonances may be present in other devices. Further investigation is required to confirm this possibility.



Moreover, the linear \textit{I}-\textit{V} curves observed for frequencies $>$1\,Hz for BFO and $>$0.05\,Hz for DBMD suggest a predominance of resistive behavior due to the dominance of odd harmonics in the amplitude spectra. However, the irregular converging harmonics in the frequency spectrum could be possibly originating from capacitive effects or internal chaotic behavior. At higher frequencies, the capacitive effects within devices become more pronounced, resulting in a capacitor-resistor-capacitor structure. In BFO, the nearly symmetric \textit{I}-\textit{V} curves result in diminished or nullified even-order terms in the signal's Taylor series expansion. As a result, odd-order harmonics dominate the spectrum, despite the presence of nonlinearities. However, for DBMD, the high-frequency \textit{I}-\textit{V} curves exhibit slightly more symmetry compared to the low-frequency curves. This leads to non-uniform convergence of harmonics, with irregular dominance of even or odd harmonics as frequency increases. For higher frequencies of more than 10\,Hz, the prevalence of odd harmonics becomes more apparent. This suggests a clear trend towards the dominance of odd harmonics. Therefore, the devices can be characterized as displaying linear resistor-like behavior at high frequencies and pronounced inherent nonlinear characteristics at low frequencies. Further investigation through experiments is needed to clarify the specific mechanisms contributing to these observed behaviors and their implications for device performance.

The phase plot of frequency spectra of BFO and DBMD are shown in Fig.~\ref{fig:4}(c) and Fig.~\ref{fig:5}(c), respectively. At low frequencies, due to a balance between multiple frequencies, components such as resistive, capacitive, and inertia effects contribute to phase shifts ranging between $0^{\circ}$ to $360^{\circ}$. The wide range of phase angles, and not just the multiples of 90, suggests the mixing of even and odd harmonics at low frequencies, as observed in the amplitude spectrum. A more or less uniform and continuous spiral pattern indicates that the system exhibits nonlinear behavior with respect to phase shifts. Moreover, at higher frequencies, the phase shifts are limited to multiples of $\pm90^{\circ}$, which could be due to the reduced capacitive reactance and increased inertia effects or inductive reactance. This results in phase shifts of even (odd) harmonics at $\pm180^{\circ}$ and odd (even) harmonics at $\pm90^{\circ}$ degrees for BFO (DBMD). From these observations, we believe that the intricate combination of device-specific characteristics, nonlinear effects, impedance dynamics, and material factors contributes to the distinct phase shift patterns observed in the memristive devices that are different from one another. For more insights, frequency spectrum plots at different $V_{\rm Device, max}$ are shown in Fig. S3 and Fig. S4 of the supplementary material.



To investigate the complex dynamics of memristive devices at various frequencies further, we conducted a detailed analysis of the phase-space attractor plots for both the BFO and DBMD devices. As shown in Fig.~\ref{fig:4}(d) for the BFO device and Fig.~\ref{fig:5}(d) for the DBMD device, we observed unique trajectories corresponding to five consecutive cycles of sinusoidal voltages. At lower frequencies, the phase-space trajectory quickly transitions from the initial conditions into a relatively stable orbit, indicating little or no chaotic behavior. However, as the frequency increases, we observe a sequence of trajectories typically associated with changes in initial conditions. Memristive devices exhibit a shift in internal state with each cycle due to the reduced response time of the particles at higher frequencies, which prevents them from returning to their original positions. This leads to a change in initial conditions for every cycle, and therefore we observe a horizontal shift in the attractors. At even higher frequencies, the device tends to exhibit linear \textit{I}-\textit{V} curves, and the attractors display an absence of any noticeable pattern, suggesting a more chaotic behavior.

            \begin{figure}[!t]
                \centering      \includegraphics[width=0.85\textwidth]{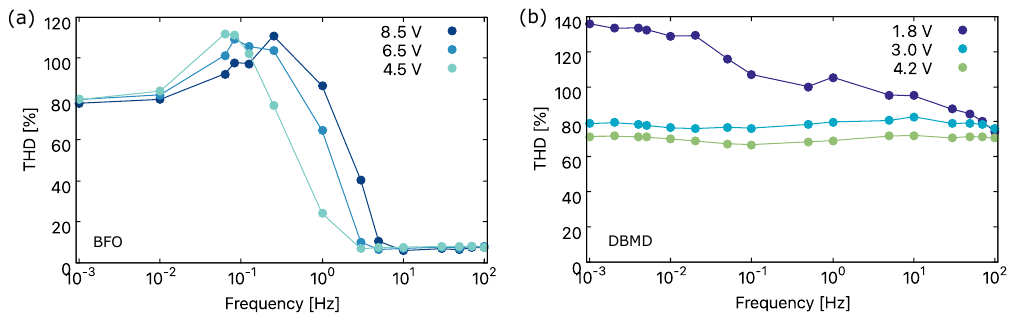}
                \caption{Total harmonic distortion in (a) BFO and (b) DBMD memristive device}
                \label{fig:6}
            \end{figure}
            
At high frequencies, the vacancies within the device may not have sufficient time to respond, resulting in a loss of pattern in the internal state and leading to chaotic behavior that makes the device's response unpredictable. However, we observed that the overall response of the device still exhibited linear characteristics, particularly when the input signals were within a certain frequency range (> 10\,Hz). In summary, although the device may display chaotic behavior at the microscopic level, the behavior may still appear linear at the macroscopic level. It is important to note that the presence of chaotic behavior does not necessarily imply nonlinearity at the macroscopic level. Chaotic behavior can arise from nonlinear dynamics, but it can also occur in linear systems under certain conditions. The findings emphasize the interplay between chaotic and linear behavior in memristive devices and the significance of analyzing both aspects for performance evaluation.


Furthermore, by subjecting memristive devices to a known input signal and analyzing the resulting output signal, one can calculate the total harmonic distortion (THD) as a measure of the distortion introduced by harmonics. Specific THD signatures based on the inherent nonlinear characteristics of different memristive devices or materials can provide a unique identifier. Variations in THD values can indicate differences in the structural or material properties of the memristive device, which contribute to its unique fingerprint. Using THD as a unique identifier for memristive devices enhances identification by introducing greater specificity and complexity to complement other characterization methods, such as \textit{I}-\textit{V} curves and frequency spectrum analysis. 

At lower frequencies, the interactions between resistive, capacitive, and inertia components are more, resulting in gradually reducing harmonics and higher. The higher THD is observed for both BFO and DBMD devices, as seen in Fig.~\ref{fig:6}. This is because, at lower frequencies, the reactive components (capacitive and inductive) have more time to affect the signal, leading to phase shifts and distortion. In other words, as mentioned above, there is an equal contribution from both even and odd harmonics. On the other hand, higher frequencies exhibit reduced nonlinear effects, and the interactions between different elements tend to be less complex compared to lower frequencies. The reactive components have less time to affect the signal, resulting in a nearly linear resistor-like response with lower THD. 

The preceding discussion indicates that the frequency spectra of memristive devices can function as a fingerprint, similar to the well-established \textit{I}-\textit{V} curves that are critical for device identification and characterization. This observation enhances our comprehension of the nonlinear behavior of these devices. Engineers can use the frequency-dependent behavior to improve device performance, gain insights into device behavior, and implement neuromorphic functionalities such as pattern recognition, memory, and learning. Fourier analysis decomposes the current response into frequency components, which can identify specific patterns within the signal, as shown in Figs. \ref{fig:4}(b)-(c) and Figs. \ref{fig:5}(b)-(c). These features can be used for machine learning algorithms to classify different patterns.

The frequency response of memristive devices can also be applied in the hardware security domain to detect tampering. Any discrepancies in the response would indicate that the device has been tampered with. Additionally, the frequency response of a memristive device can be used to generate random numbers. This is possible because the frequency response of a memristive device is inherently nonlinear and exhibits chaotic behavior, as shown in Fig. \ref{fig:4}(d) and Fig. \ref{fig:5}(d). By applying a periodic input signal to the device and observing the output signal, which is the device's response, one can derive a sequence of random numbers. The chaotic nature of the device ensures that the generated sequence of random numbers is unpredictable and statistically random. A detailed analysis of the statistical properties, e.g., true randomness, remains for future work.


\section*{Conclusion}

This paper explores the nonlinear behavior of BFO and DBMD devices using Fourier analysis, which is crucial for applications such as neuromorphic computing and hardware security. A physics-based compact model based on the cloud-in-a-cell scheme is used to investigate the simultaneous presence of resistive, capacitive, and inductive effects. The capacitance arises from the change in depletion layer width at the Schottky contacts, tunnel barrier, and within the oxide due to the movement of ions or vacancies. Inductive effects occur due to the inertia of the movable defects. The \textit{I}-\textit{V} curves show that capacitive effects have a significant impact on the resistive switching of the BFO device, resulting in non-zero crossing hysteresis. In contrast, DBMD's behavior is less affected due to its lower capacitance. The frequency-dependent \textit{I}-\textit{V} curves of both devices exhibit decreasing hysteresis at higher frequencies, approaching linear resistor-like behavior. However, the amplitude spectrum reveals increased nonlinearity at low frequencies and uniform convergence of harmonics. Meanwhile, for high frequencies, the dominance of odd harmonics indicates reduced nonlinear behavior but increased chaotic behavior observed from the internal state attractors due to reduced time for the particles to react. 
The corresponding phase plots show spiral patterns at low frequencies with randomly varying angles indicating the mixing of different harmonics, while for high frequencies, the phase changes only as multiples of 90 degrees. 
Based on the observations from this study, we propose using frequency spectra as a fingerprint in addition to conventional \textit{I}-\textit{V} curves to better understand nonlinear dynamics.

\section*{Methods}
\noindent\textbf{Simulations} An in-house model developed using C programming language.  The codes developed and implemented are currently research codes that are not yet ready to be used by non-experts.

\section*{Data Availability Statement}
The datasets used and/or analysed during the current study available from the corresponding author on reasonable request.

\bibliography{sample}

\section*{Acknowledgements}
Funded by the Deutsche Forschungsgemeinschaft (DFG, German Research Foundation) - Project-ID 434434223- SFB 1461 and Project-IDs 439700144 and 440182124 - Research Grants MU 2332/10-1 and AR 1387/1-1 in the frame of Priority Program SPP 2253.

\section*{Author contributions statement}
S.Y. prepared the data sets, implemented the methodology, conducted the simulations, analyzed the results, and prepared the manuscript. T.G. and J.T. provided critical feedback, were involved in discussions, and helped shape the research. F.J. reviewed the manuscript and was involved in discussions. T.M. and T.A. conceived and directed the conceptual ideas of the work. S. Y., T.H., and T.M. developed the methodological concept for memristor modeling. T.M. acquired the funding and administered the project. All authors reviewed the manuscript and contributed to the editing.

\section*{Competing interests}
The authors declare no competing interests.

\end{document}


\section*{Supplementary material}

\vspace{0.5cm}
\noindent
{\large \textbf{Nonlinear behavior of memristive devices for hardware \\
security primitives and neuromorphic computing systems}}

\vspace{1cm}
\subsection*{Flowchart of the proposed model:}
            \renewcommand{\thefigure}{S1} 
            \begin{figure}[!ht]
                \centering             \includegraphics[width=0.9\textwidth]{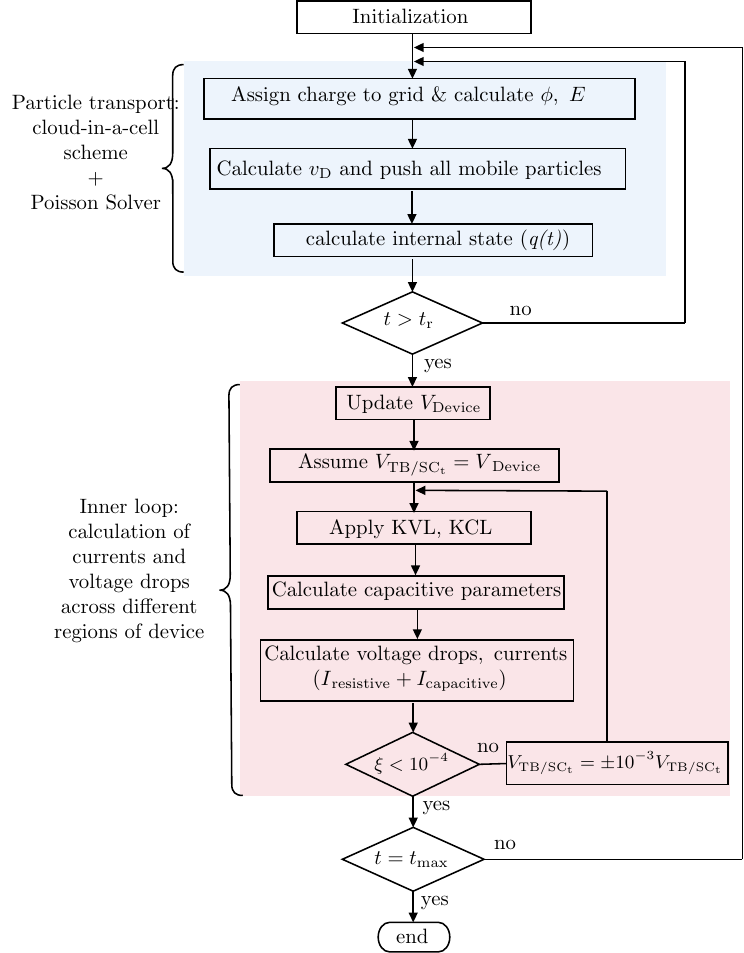}
                \caption{Flowchart describing all the steps involved in the proposed model}
                \label{fig:flowchart}
            \end{figure}

\newpage
\vspace{1cm}
\subsection*{Cloud-in-a-cell approach:}
            \renewcommand{\thefigure}{S2} 
            \begin{figure}[!ht]
                \centering             \includegraphics[width=1.0\textwidth]{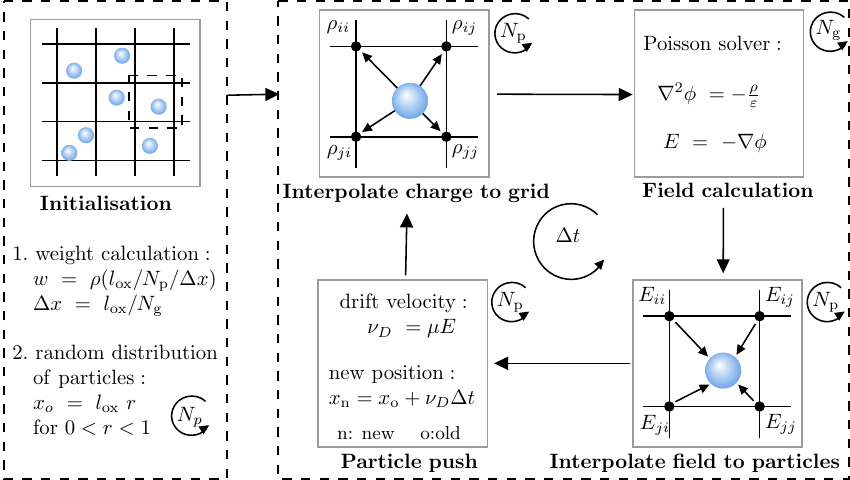}
                \caption{Flowchart describing the steps involved in cloud-in-a-cell (CIC) appraoch}
                \label{fig:3_CIC}
            \end{figure}

The cloud-in-a-cell approach is a specific interpolation scheme that falls under the broader category of methods known as particle-in-cell (PIC) simulations. The CIC approach is used to study the behavior of particles (charged defects) in a metal oxide that contributes to resistive switching in RRAMs. In this approach, the RRAM device is discretized into a grid of cells ($N_{\mathrm{g}}$) resembling a crystal lattice. Each cell represents a small volume within the device. The particles, representing atoms, ions, vacancies, or electrons, are conceptualized as forming a "cloud" around their actual positions. To distribute the properties of these particles onto the grid, the cloud around each particle is spread over multiple adjacent grid points. The initial position of each particle is given by: 

\begin{equation}
    x_{o} = l_{\mathrm{ox}} \, r, \tag{S1} 
    \label{Eq:3.22}
\end{equation}

\noindent where $r$ is any random number between 0 and 1 and $l_{\mathrm{ox}}$ is the length of the oxide. The contributions of $N_{\mathrm{p}}$ particles are weighted based on their positions within the grid cells. In other words, particles are given weights based on the defect density ($\rho$) of the oxide. Here, weight refers to a value assigned to each simulated particle. The weight represents the number of real particles that the simulated particle represents. In CIC simulations, it is not practical to simulate every particle in an oxide layer, due to computational limitations. Instead, a small number of simulated particles are used to represent a large number of real particles. The weight of each simulated particle is determined based on the ratio of the total number of real particles in the oxide layer to the number of simulated particles. Mathematically, it can be represented by the below formula:

\begin{equation}
    w = \rho \frac{l_{\mathrm{ox}}\Delta x}{N_{\mathrm{p}}}, \tag{S2}
    \label{Eq:3.23}
\end{equation}

with $\Delta x$ as the distance between two grid points. Then, one critical aspect of the CIC approach is how the charge density at each grid point is calculated. The charge carried by particles influences the electric field within the device. To calculate the charge density, the particles' charges are distributed to the grid cells based on their positions. This process involves linear interpolation techniques, where the charge from each particle is distributed among neighboring cells according to their position and weight. The charge density is calculated as follows:

\begin{equation}
    \rho_{i/i+1} = \rho_{o,i/i+1} + eDw, \tag{S3}
    \label{Eq:3.24}
\end{equation}

where, $D$ is the distance of the particles from their position to grid cells $i$ or $i+1$, $\rho_{o,i/i+1}$ is the already present charge density (if any), and e is the elementary charge. This interpolated charge density is then used to update the electric field in the device. The simulation proceeds in discrete time steps. In each time step, the particles' positions and velocities are updated based on the forces they experience, such as electric fields. The complete process is further illustrated in Fig. \ref{fig:3_CIC}.

\vspace{1cm}
\subsection*{Poisson Solver:}

 In many cases, particularly in this work, for the analysis of memristive devices, the 1D Poisson solver serves as a fundamental tool for calculating the distribution of electric potential and electric field in the oxide layer. 

The 1D Poisson equation is derived from Gauss's law, which states that the divergence of the electric field in a region of space is proportional to the charge density within that region. Mathematically, it can be represented as
\begin{equation}
    \frac{d}{dx} \left ( \varepsilon \frac{d\phi}{dx}\right ) = -\rho, \tag{S4}
    \label{Eq:3.18}\end{equation}
\noindent where $\varepsilon $ is the permittivity of the oxide layer and $\rho$ is the charge density. The potential equation is subjected to Dirichlet boundary conditions at the simulation domain boundaries of the electrodes. For boundaries other than electrodes, periodic boundary conditions are applied. The Coulomb interactions between different particles are not taken into account. The electric field is then by 
\begin{equation}
E = -\frac{d\phi}{dx}. \tag{S5}
\label{Eq:3.19}
\end{equation}
To numerically compute the electric potential and electric field, Eq.\eqref{Eq:3.18} and Eq.\eqref{Eq:3.18} are discretized into a 1D mesh. The successive over-relaxation (SOR) method iteratively updates the solution to the Poisson equation. Eq.\eqref{Eq:3.18} along the x-axis can be written as,
\begin{equation}
     \phi_{i}^{t} =  0.5 * \frac{\rho \Delta x^{2}}{\varepsilon} + 0.5\left ( \phi_{i+1}^{t} + \phi_{i-1}^{t} \right ), \tag{S6}
     \label{Eq:3.20}
\end{equation}
\begin{equation}
    \phi_{i}^{t} =  (1-\omega_{\textrm{SOR}})\phi_{i}^{t-1} + \omega_{\textrm{SOR}}\phi_{i}^{t}. \tag{S7}
    \label{Eq:3.21}
\end{equation}

In this equation, the relaxation parameter,
\begin{equation}
     \omega_{\textrm{SOR}} = \frac{2}{1+\sqrt{1-cos^{2}\left ( \frac{\pi }{N_{\textrm{g}}} \right )}}. \tag{S8}
     \label{Eq:3.25}
\end{equation}
The electric field in the discretized form can be written as
\begin{equation}
    E_{i}^{t} = =\left (\frac{\phi_{i}^{t} - \phi_{i-1}^{t}}{2\Delta x} \right ). \tag{S9}
    \label{Eq:3.26}
\end{equation}

\newpage
\vspace{1cm}
\subsection*{Device parameters:}

The BFO device parameters used to obtain plots in Fig.~3(a)-(d), Fig.~4(a)-(b), Fig.~5 and Fig.~7(a).

\begin{table}[!ht]
\centering
\begin{tabular}{|l|c|c|}
\hline
\textbf{Physical quantity} & \textbf{Symbol} & \textbf{Value} \\
\hline
Temperature & $T$ & 298\,K \\
\hline
Phonon frequency & $\nu_{0}$ & $1 \times 10^{12}$\,Hz\\
\hline
Lattice constant & $d$ & $0.56 \times 10^{-9}$\,m\\
\hline
Device area & $A_{\rm d}$ & $0.04$\,$\rm mm^{2}$\\
\hline
Relative permittivity of $\rm BiFeO_{3}$ & $\varepsilon_{r} $ & 52.0\\
\hline
Activation energy de-trapping & ${ U}_{\rm A}$ & 0.55\,eV\\
\hline
Activation energy trapping & ${ U}_{\rm A}$ & 0.76\,eV\\
\hline
Conductivity of $\rm BiFeO_{3}$  & $\sigma$ & $7.0 \times
 10^{-4}$\,$\Omega {\rm m}$\\
 \hline
 Length of BFO & $l_{\rm BFO}$ & $600 \times 10^{-9}$\,m\\
 \hline
Defect density  & $\rho$ & $8\times 10^{16}$\,${\rm cm^{-3}}$ \\
\hline
Top Schottky contact height & $\Phi_{\rm SC_{t}}$ & $0.75$\,eV\\
\hline
Top Schottky contact ideality factor & $n_{\rm SC_{t}}$ & 4.0\\
\hline
Fitting parameter top SC depletion layer & $\lambda_{\rm d_{t}}$ & 0.3\\
\hline
Bottom Schottky contact height  & $\Phi_{\rm SC_{b}}$ & $0.85$\,eV\\
\hline
Bottom Schottky contact ideality factor & $n_{\rm SC_{b}}$ & 4.42\\
\hline
Fitting parameter bottom SC depletion layer & $\lambda_{\rm d_{b}}$ & -0.9\\
\hline
\end{tabular}
\caption*{Table S1: Details of the simulation parameters of BFO device.}
\label{tab:S1}
\end{table}

The DBMD device parameters used to obtain plots in Fig.~3(e)-(h), Fig.~4(c)-(d), Fig.~6 and Fig.~7(b).
\begin{table}[!ht]
\centering
\begin{tabular}{|l|c|c|}
\hline
\textbf{Physical quantity} & \textbf{Symbol} & \textbf{Value} \\
\hline
Temperature & $T$ & 298\,K \\
\hline
Lattice constant & $d$ & $2.5 \times 10^{-10}\, {\rm m}$\\
\hline
Device area & $A_{\rm d}$ & $625\, \mu{\rm {m}^{2}}$\\
\hline
Relative permittivity ($\rm Nb_{x}O_{y}$) & $\varepsilon_{r} $ & 42.0\\
\hline
Activation energy & ${ U}_{\rm A}$ & 0.76\,eV\\
\hline
Conductivity ($\rm Nb_{x}O_{y}$) & $\sigma$ & $1.0 \times
 10^{-4}\, \Omega {\rm m}$\\
 \hline
 Length of SE ($\rm Nb_{x}O_{y}$) & $l_{\rm SE}$ & $2.5 \times 10^{-9}\,{\rm m}$\\
 \hline
Defect density & $\rho$ & $5\times 10^{20}\, {\rm cm^{-3}} $\\
\hline
Tunnel barrier width & $d_{\rm TB}$& $1.1 \times 10^{-9}\, {\rm m}$\\
\hline
Tunnel barrier height & $\Phi_{\rm TB}$ & $3.2$\,eV\\
\hline
Fitting parameter TB depletion layer & $\lambda_{\rm t}$ & 0.1\\
\hline
Schottky contact height & $\Phi_{\rm SC}$ & $0.96$\,eV\\
\hline
Schottky contact ideality factor & $n_{\rm SC}$ & 4.2\\
\hline
Fitting parameter SC depletion layer & $\lambda_{\rm d}$ & -0.7\\
\hline
\end{tabular}
\caption*{Table S2: Details of the simulation parameters of DBMD device.}
\label{tab:S2}
\end{table}

\newpage
\subsection*{Frequency spectra of BFO similar to Fig. 5 but different voltage amplitudes}
\renewcommand{\thefigure}{S3} 
\begin{figure}[!ht]
\centering\includegraphics[width=0.59\textwidth]{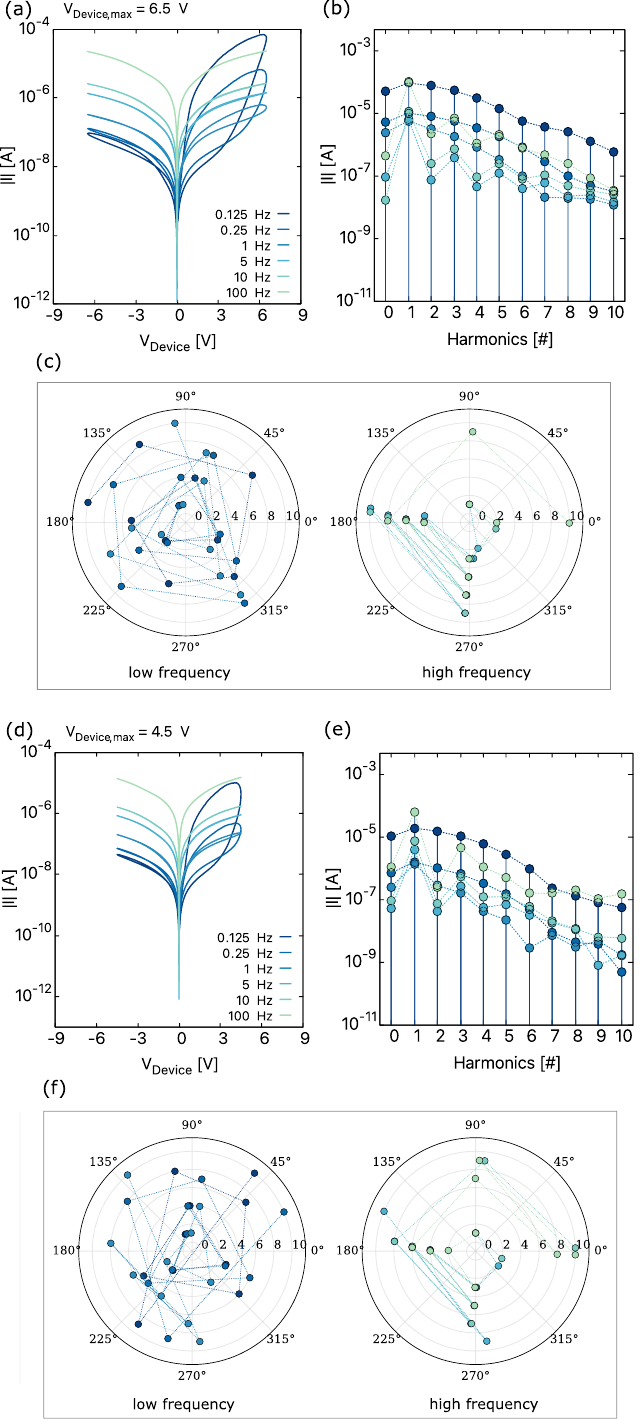}
\caption{(a),(d) The simulated current-voltage characteristics of BFO obtained for a sinusoidal input voltage of 6.5\,V and 4.5\,V with various frequencies, Frequency spectra of BFO corresponding to the \textit{I}-\textit{V} curves depicted in Figure (a) and (d) showing (b),(e) amplitude spectra and (c),(f) phase spectra.}
\label{fig:3}
\end{figure}

\newpage
\subsection*{Frequency spectra of DBMD similar to Fig. 6 but different voltage amplitudes}
\renewcommand{\thefigure}{S4} 
\begin{figure}[!ht]
\centering\includegraphics[width=0.6\textwidth]{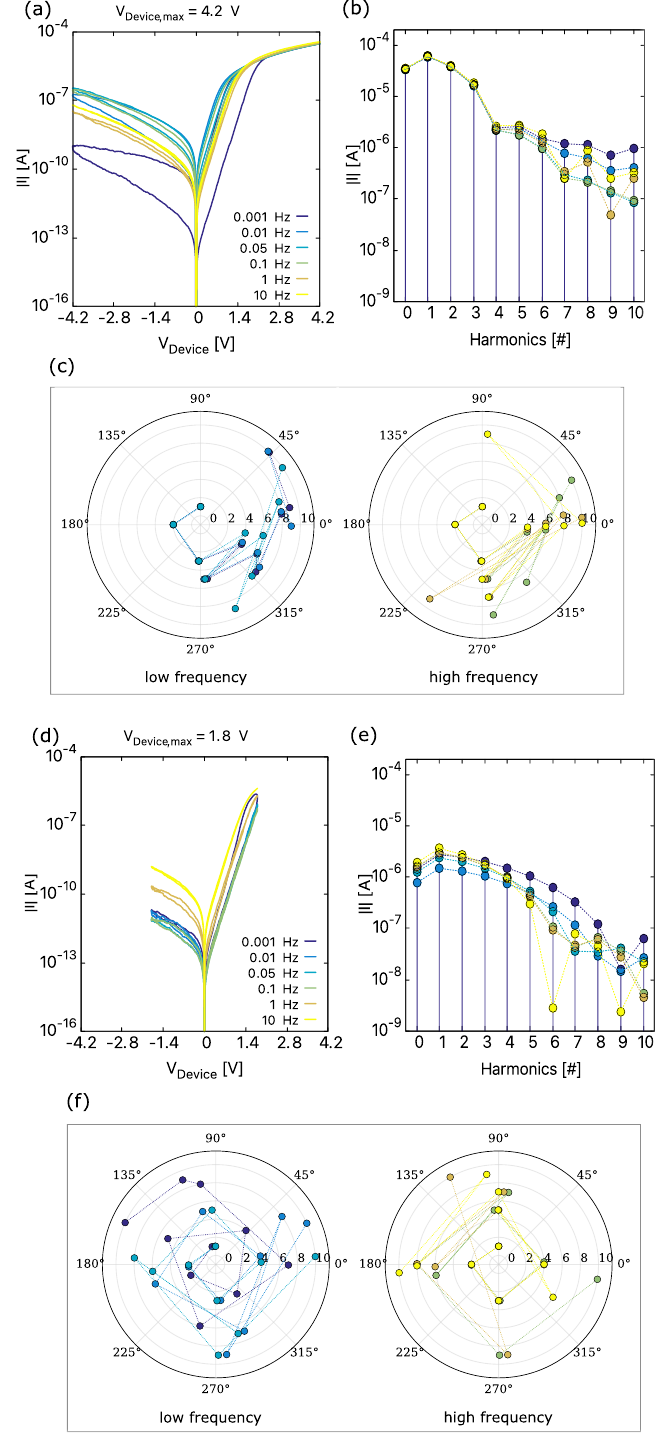}
\caption{(a),(d) The simulated current-voltage characteristics of DBMD obtained for a sinusoidal input voltage of 4.2\,V and 1.8\,V with various frequencies, Frequency spectra of DBMD corresponding to the \textit{I}-\textit{V} curves depicted in Figure (a) and (d) showing (b),(e) amplitude spectra and (c),(f) phase spectra.}
\label{fig:4}
\end{figure}